\documentclass[graybox]{svmult}

\usepackage{mathptmx}        % selects Times Roman as basic font
\usepackage{makeidx}         % allows index generation
\usepackage{cite}
\usepackage{graphicx}        % standard LaTeX graphics tool
\usepackage{grffile}
\usepackage{epstopdf}        % when including figure files
\usepackage{amsmath}
\usepackage{multicol}        % used for the two-column index
\usepackage[bottom]{footmisc}% places footnotes at page bottom

\makeindex             % used for the subject index
\begin{document}

%\title*{New families of periodic orbits for the planar
%three-body problem found by Newton's method}

\title*{Newton's method for computing  periodic orbits of the planar
three-body problem}

\author{I. Hristov$^{1,a}$, R. Hristova$^1$, I. Puzynin$^2$,  T. Puzynina$^2$, Z. Sharipov$^{2,b}$, Z. Tukhliev$^2$\\
\vspace{0.5cm}
\emph{$^1$ Faculty of Mathematics and Informatics, Sofia University, Sofia, Bulgaria}\\
\emph{$^2$ Meshcheryakov Laboratory of Information Technologies, JINR,  Dubna, Russia}\\
\vspace{0.5cm}
\emph{E-mails:  $^a$  ivanh@fmi.uni-sofia.bg \hspace{0.3cm} $^b$ zarif@jinr.ru}}
\authorrunning{I. Hristov, R. Hristova, I. Puzynin et al.}
\maketitle

\textbf{Keywords}: Periodic orbits of the planar three-body problem, Newton's method, High order multiple precision Taylor series method \\

\abstract {In this paper we present in detail Newton's method and its modification, based on the Continuous analog of  Newton's method
          for computing periodic orbits of the planar three-body problem.
          The linear system at each step of the method is formed by solving
          a system of ODEs with the multiple precision Taylor series method.
          We consider zero angular momentum symmetric initial configuration with parallel velocities,
          bodies with equal masses and relatively short periods. Taking candidates for the correction method with greater return proximity
          as usual and correcting with the modified Newton's method, allows us to find some new topological families that are not included in the database
          in [SCIENCE CHINA Physics, Mechanics \& Astronomy 60.12 (2017)].}

\section{Introduction}
A breakthrough in numerical searching of  periodic orbits of the planar three-body problem
was  made in recent years. In 2013 Shuvakov and Dmitrashinovich found 11 new topological families applying
a clever numerical algorithm in the standard double precision arithmetic \cite{Shuv1}, \cite{Shuv2}.
Since the three-body problem is very sensitive on the initial conditions,
working with double precision strongly limits the number of solutions that can be found.
This limitation was recognized by Li and Liao, in 2017 they applied Newton's method to find
more than 600 new families of periodic orbits \cite{Liao1}. They formed the linear system
at each step of  Newton's method by solving a system of ODEs with the high order multiple precision
Taylor series method. However, no details of the numerical procedure are given in \cite{Liao1}.
This numerical procedure is rather technical and deserves its own attention. In this work
we present in detail Newton's method and also its modification, based on the Continuous analog of  Newton's method \cite{CANM}
for computing periodic orbits of the planar three-body problem. Our programs are tested in relatively short periods.
We take candidates for correction with greater than usual return proximity and correct them with
the modified Newton's method. As a result we find some new topological families that
are not included in the database in  \cite{Liao1}.

It is important to mention that the high order multiple precision Taylor series method can be
applied in a very promising numerical procedure, with lots of applications, called
Clean Numerical Simulation \cite{Liao2},\cite{Liao3}.

\section{The mathematical model}

The differential equations for the three-body problem are derived
from Newton's second law and Newton's law of gravity:
$$
m_i\ddot{r}_i=\sum_{j=1,j\neq i}^{3}G m_i m_j \frac{(r_j-r_i)}{{\|r_i -r_j\|}^3}, i=1,2,3.
$$
where $G$ is the gravitational constant and $m_i$ are the masses of the bodies.
In this paper we consider normalized equations with $G=1$ and bodies with equal masses
$m_1=m_2=m_3=1$. The model considers the bodies as mass points. We solve the following system of equations:
\begin{equation}
\ddot{r}_i=\sum_{j=1,j\neq i}^{3} \frac{(r_j-r_i)}{{\|r_i -r_j\|}^3}, i=1,2,3.
\end{equation}
Planar motion of the three bodies is considered, so the vectors $r_i$, $\dot{r}_i$ have two components: $r_i=(x_i, y_i)$,
$\dot{r}_i=(\dot{x}_i, \dot{y}_i).$
The variables ${vx}_i$ and ${vy}_i$ are introduced, so that ${vx}_i=\dot{x}_i, {vy}_i=\dot{y}_i.$
Then the second order system (1) can be written as a first order one:
\begin{equation}
\dot{x}_i={vx}_i, \hspace{0.1 cm} \dot{y}_i={vy}_i,  \hspace{0.1 cm} \dot{vx}_i=\sum_{j=1,j\neq i}^{3} \frac{(x_j-x_i)}{{\|r_i -r_j\|}^3},  \hspace{0.1 cm} \dot{vy}_i=\sum_{j=1,j\neq i}^{3} \frac{(y_j-y_i)}{{\|r_i -r_j\|}^3}, \hspace{0.1 cm} i=1,2,3
\end{equation}
The system is solved numerically in this first order form. So we have a vector of 12 unknown functions
$ u(t)={(x_1, y_1, {vx}_1, {vy}_1, x_2, y_2, {vx}_2, {vy}_2, x_3, y_3, {vx}_3, {vy}_3)}^\top$.
We search for periodic planar collisionless orbits as in \cite{Shuv1}, \cite{Shuv2}, \cite{Liao1}: with zero angular momentum and symmetric initial
configuration with parallel velocities (Fig. 1):
$$(x_1(0),y_1(0))=(-1,0), \hspace{0.2 cm} (x_2(0),y_2(0))=(1,0), \hspace{0.2 cm} (x_3(0),y_3(0))=(0,0)$$
$$({vx}_1(0),{vy}_1(0))=({vx}_2(0),{vy}_2(0))=(v_x,v_y)$$
$$({vx}_3(0),{vy}_3(0))=-2({vx}_1(0),{vy}_1(0))=(-2v_x, -2v_y),$$
where $v_x\in [0,1], v_y\in [0,1]$ are parameters. Let us denote the periods of the orbits with $T$. Our goal is to find triplets $(v_x, v_y, T)$
for which the periodicity condition $u(T)=u(0)$ is true. In what follows we will use the same notation for $v_x, v_y, T$
and their approximations.
\begin{figure}
\begin{center}
\includegraphics[scale=0.3]{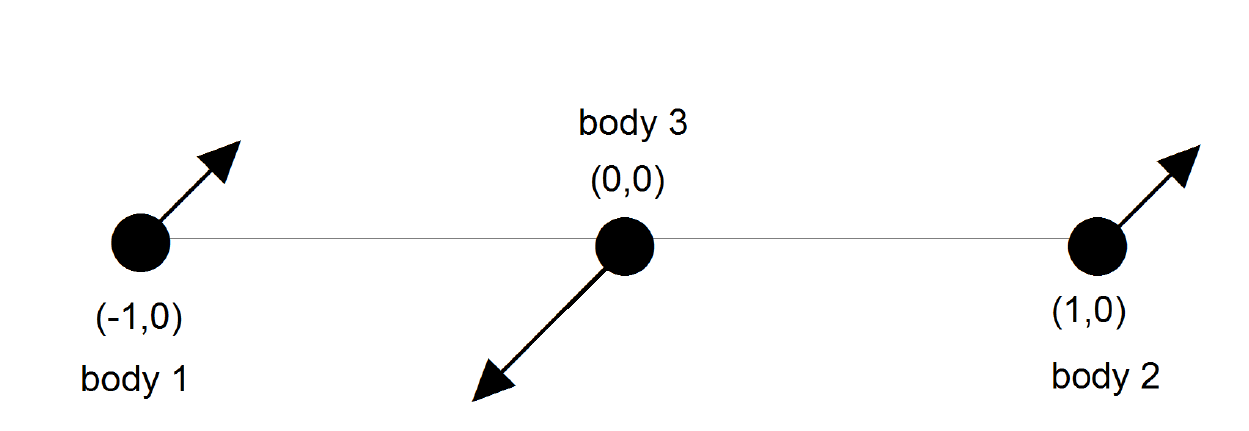}
\caption{Initial configuration}
\label{fig:1}
\end{center}
\end{figure}
\section{Short description of the numerical procedure}
The numerical procedure for searching periodic orbits consists of two stages: stage (I) - computing initial approximations
for the correction method, stage (II) - applying correction method, which can converge or diverge.
We use the grid-search algorithm at stage (I) for computing initial approximations for the correction method.
We use a $1024 \times 1024$ grid in the square $[0,1]\times[0,1]$ for
parameters $v_x$, $v_y$. We consider relatively short periods $T$: $T\leq50$. This condition does not exclude to find solutions with periods slightly greater then 50,
when we use the correction method.
Let $$P(t)=\sqrt{\sum_{i=1}^{3}{\|r_i(t)-r_i(0)\|}^{2}+\sum_{i=1}^{3}{\|\dot{r}_i(t)-\dot{r}_i(0)\|}^{2}}$$ be the proximity function.
As candidates for the correction method, we take those triplets  $(v_x, v_y, T)$
for which the return proximity is less then $0.7$:
$$P(T) = \min_{1<t\leq50}P(t) < 0.7$$
The number 0.7 is greater than the criterion numbers that are usually used.
We also set the condition that the return proximity is a local minimum on the grid for $v_x, v_y$.
As a correction method  we use  a modification of Newton's method based on continuous  analog of Newton's method  \cite{CANM}.
Our correction method has a larger domain of convergence than the classic Newton's method. This fact, together with the relaxed
condition for the return proximity, allows us to find some new topological families for $T\leq50$ that do not exist in the database
from \cite{Liao1}.

\section{The linear system at each step of Newton's method}
\label{sec:2}
We follow the general description of Newton's method for computing periodic orbits in \cite{Barrio3}.
Let  $v_x, v_y, T$ are approximations of the initial velocities and the period
for some periodic solution: $u(T)\approx u(0)$.
These approximations are improved with corrections $\Delta v_x,
\Delta v_y, \Delta T$ by expanding the periodicity condition
in a multivariable Taylor series up to the first order:

$$
\begin{pmatrix}
x_1(T) \\
y_1(T) \\
{vx}_1(T)\\
{vy}_1(T)\\
x_2(T) \\
y_2(T) \\
{vx}_2(T)\\
{vy}_2(T)\\
x_3(T) \\
y_3(T) \\
{vx}_3(T)\\
{vy}_3(T)
\end{pmatrix}
+
\begin{pmatrix}
\frac{\partial x_1}{\partial v_x}(T) & \frac{\partial x_1}{\partial v_y}(T) & \dot{x}_1(T)\\
\frac{\partial y_1}{\partial v_x}(T) & \frac{\partial y_1}{\partial v_y}(T) & \dot{y}_1(T)\\
\frac{\partial {vx}_1}{\partial v_x}(T)& \frac{\partial {vx}_1}{\partial v_y}(T) & \dot{vx}_1(T)\\
\frac{\partial {vy}_1}{\partial v_x}(T) & \frac{\partial {vy}_1}{\partial v_y}(T)& \dot{vy}_1(T)\\
\frac{\partial x_2}{\partial v_x}(T) & \frac{\partial x_2}{\partial v_y}(T) & \dot{x}_2(T)\\
\frac{\partial y_2}{\partial v_x}(T) & \frac{\partial y_2}{\partial v_y}(T) & \dot{y}_2(T)\\
\frac{\partial {vx}_2}{\partial v_x}(T) & \frac{\partial {vx}_2}{\partial v_y}(T) & \dot{vx}_2(T)\\
\frac{\partial {vy}_2}{\partial v_x}(T) & \frac{\partial {vy}_2}{\partial v_y}(T) & \dot{vy}_2(T)\\
\frac{\partial x_3}{\partial v_x}(T) & \frac{\partial x_3}{\partial v_y}(T) & \dot{x}_3(T)\\
\frac{\partial y_3}{\partial v_x}(T) & \frac{\partial y_3}{\partial v_y}(T) & \dot{y}_3(T)\\
\frac{\partial {vx}_3}{\partial v_x}(T) & \frac{\partial {vx}_3}{\partial v_y}(T) & \dot{vx}_3(T)\\
\frac{\partial {vy}_3}{\partial v_x}(T) & \frac{\partial {vy}_3}{\partial v_y}(T) & \dot{vy}_3(T)
\end{pmatrix}
\begin{pmatrix}
\Delta v_x \\
\Delta v_y \\
\Delta T
\end{pmatrix}
=
\begin{pmatrix}
x_1(0) \\
y_1(0)  \\
{vx}_1(0) + \Delta v_x\\
{vy}_1(0) + \Delta v_y\\
x_2(0) \\
y_2(0) \\
{vx}_2(0) + \Delta v_x\\
{vy}_2(0) + \Delta v_y\\
x_3(0)  \\
y_3(0)  \\
{vx}_3(0)- 2 \Delta v_x\\
{vy}_3(0)- 2 \Delta v_y
\end{pmatrix}
$$
By reordering this equation we obtain the following linear system with a $12\times 3$ matrix with respect to ${(\Delta v_x, \Delta v_y, \Delta T)}^\top$.
\begin{equation}
\begin{pmatrix}
\frac{\partial x_1}{\partial v_x}(T) & \frac{\partial x_1}{\partial v_y}(T) & \dot{x}_1(T)\\
\frac{\partial y_1}{\partial v_x}(T) & \frac{\partial y_1}{\partial v_y}(T) & \dot{y}_1(T)\\
\frac{\partial {vx}_1}{\partial v_x}(T) -1 & \frac{\partial {vx}_1}{\partial v_y}(T) & \dot{vx}_1(T)\\
\frac{\partial {vy}_1}{\partial v_x}(T) & \frac{\partial {vy}_1}{\partial v_y}(T) - 1 & \dot{vy}_1(T)\\
\frac{\partial x_2}{\partial v_x}(T) & \frac{\partial x_2}{\partial v_y}(T) & \dot{x}_2(T)\\
\frac{\partial y_2}{\partial v_x}(T) & \frac{\partial y_2}{\partial v_y}(T) & \dot{y}_2(T)\\
\frac{\partial {vx}_2}{\partial v_x}(T) -1 & \frac{\partial {vx}_2}{\partial v_y}(T) & \dot{vx}_2(T)\\
\frac{\partial {vy}_2}{\partial v_x}(T) & \frac{\partial {vy}_2}{\partial v_y}(T) - 1 & \dot{vy}_2(T)\\
\frac{\partial x_3}{\partial v_x}(T) & \frac{\partial x_3}{\partial v_y}(T) & \dot{x}_3(T)\\
\frac{\partial y_3}{\partial v_x}(T) & \frac{\partial y_3}{\partial v_y}(T) & \dot{y}_3(T)\\
\frac{\partial {vx}_3}{\partial v_x}(T) +2 & \frac{\partial {vx}_3}{\partial v_y}(T) & \dot{vx}_3(T)\\
\frac{\partial {vy}_3}{\partial v_x}(T) & \frac{\partial {vy}_3}{\partial v_y}(T) +2 & \dot{vy}_3(T)
\end{pmatrix}
\begin{pmatrix}
\Delta v_x \\
\Delta v_y \\
\Delta T
\end{pmatrix}
=
\begin{pmatrix}
x_1(0) - x_1(T) \\
y_1(0) - y_1(T) \\
{vx}_1(0) - {vx}_1(T)\\
{vy}_1(0) - {vy}_1(T)\\
x_2(0) - x_2(T) \\
y_2(0) - y_2(T) \\
{vx}_2(0) - {vx}_2(T)\\
{vy}_2(0) - {vy}_2(T)\\
x_3(0) - x_3(T) \\
y_3(0) - y_3(T) \\
{vx}_3(0) - {vx}_3(T)\\
{vy}_3(0) - {vy}_3(T)
\end{pmatrix}
\end{equation}
We solve system (3) as a linear least square problem using $QR$ decomposition based on Householder reflections
\cite{Demmel}. Then we correct and obtain the next approximation:
$$v_x := v_x + \Delta v_x,\hspace{0.2 cm} v_y := v_y + \Delta v_y, \hspace{0.2 cm} T := T + \Delta T$$
We iterate until the value of the return proximity $P(T)$ at some iteration is less than some tolerance or the number of iterations become
greater than some number $maxiter$ to detect divergence. This is the classic Newton's method. In this paper
we use a modification of Newton's method based on continuous  analog of Newton's method  \cite{CANM}. We introduce a parameter
$0 < \tau_k <=1$, where k is the number of the iteration.
Now we correct this way:
$$v_x := v_x + \tau_k\Delta v_x, \hspace{0.2 cm} v_y := v_y + \tau_k\Delta v_y, \hspace{0.2 cm} T := T + \tau_k\Delta T$$
Let $P_k$ be the value of the return proximity $P(T)$ at the k-th iteration.
With given $\tau_0$ the next $\tau_k, k=1,2,...$ is computed with the following adaptive algorithm \cite{CANM}:

\begin{equation}
\tau_k = \left \{
               \begin{array}{ll}
                \min(1,\ \tau_{k-1} P_{k-1} / P_k), &
                 P_k \leq P_{k-1}, \\\\
                \max(\tau_0,\ \tau_{k-1} P_{k-1} / P_k), &
                 P_k > P_{k-1},
               \end{array}
                                  \right .
\end{equation}
The modification of  Newton's method does not add any technical difficulties, because we have to solve the same linear system.
The important thing is that this method has a larger domain of convergence and makes it possible to find
more periodic orbits for a given grid for the parameters $v_x,v_y$.
When $\tau_k=1$ for all $k$, the method matches with the classic Newton's method.
In this work we take  $\tau_0=0.2$.

\section{Computing the coefficients for the linear system with Taylor series method}
To compute the coefficients in the $12\times3$ matrix
of system (3), we have to add to the equations in (2) the 24 differential  equations for the
partial derivatives with respect to the parameters $v_x, v_y$:
$$\frac{\partial x_i}{\partial v_x}(t),\frac{\partial y_i}{\partial v_x}(t),
\frac{\partial {vx}_i}{\partial v_x}(t),\frac{\partial {vy}_i}{\partial v_x}(t),
\frac{\partial x_i}{\partial v_y}(t),\frac{\partial y_i}{\partial v_y}(t),
\frac{\partial {vx}_i}{\partial v_y}(t),\frac{\partial {vy}_i}{\partial v_y}(t), i=1,2,3.$$
These equations can be obtained by differentiation of system (2) with respect to the parameters $v_x, v_y$,
but we do not need them in explicit form.
However, we need the initial conditions and they are:
$$\frac{\partial x_i}{\partial v_x}(0)=\frac{\partial y_i}{\partial v_x}(0)=\frac{\partial x_i}{\partial v_y}(0)=\frac{\partial y_i}{\partial v_y}(0)=0, i=1,2,3$$
$$\frac{\partial {vy}_i}{\partial v_x}(0)=0, i=1,2,3, \frac{\partial {vx}_1}{\partial v_x}(0)=\frac{\partial {vx}_2}{\partial v_x}(0)=1, \frac{\partial {vx}_3}{\partial v_x}(0)=-2$$
$$\frac{\partial {vx}_i}{\partial v_y}(0)=0, i=1,2,3, \frac{\partial {vy}_1}{\partial v_y}(0)=\frac{\partial {vy}_2}{\partial v_y}(0)=1, \frac{\partial {vy}_3}{\partial v_y}(0)=-2$$
At each step of Newton's method we have to solve a system of 36 ODEs in the interval $[0,T]$ with initial conditions corresponding to $v_x, v_y.$
12 ODEs comes from system (2) and 24 ODEs comes from the partial derivatives with respect to $v_x, v_y.$
A crucial decision for the success of finding periodic orbits is the choice of the numerical algorithm for solving this system of 36 ODEs.
To follow the trajectories accurately for a long time, we use the high order multiple precision Taylor series method \cite{Jorba}, \cite{Barrio1}.
We use this method at the stage (I) also, when we compute the candidates for the correction method by solving system (2) at each grid point for $v_x$, $v_y$.
The N-th order Taylor series method for unknown vector $U(t)$ is given by the expression:
$$U(t+\tau)=\sum_{i=0}^{N} U^{[i]}\tau^i, \hspace{0.3 cm} U^{[i]}=\frac{1}{i!}\frac{dU^{(i)}(t)}{dt^i},$$
where $U^{[i]}$ are the so called normalized derivatives. The use of an adaptive step-size strategy is also crucial
for the three-body problem. The stepsize $\tau$ is determined by the last two terms of the Taylor expansions \cite{Jorba}:
$$\tau = \frac{e^{-0.7/(N-1)}}{e^{2}} \min\left\{   {\left(\frac{1}{{\|\mathbf{U^{[N-1]}} \|}_{\infty}}\right)}^{\frac{1}{N-1}}, {\left(\frac{1}{{\|\mathbf{U^{[N]}}\|}}_{\infty}\right)}^{\frac{1}{N}}\right\}$$
At stage (I) the vector $U$ has 12 components and
at stage (II) - 36 components. When integrating the systems of ODEs in $[0,T]$ we set some upper limit of the number of time-steps.

The most technical part of our algorithm is computing the normalized derivatives. They are computed by the rules
of automatic differentiation \cite{Barrio1},\cite{Barrio2}. We need to write system (2) in a form for applying the rules of automatic differentiation for sum, product of two functions and the rule for power of a function.
Let us denote with $S_{ij}$, $SS_{ij}$, $i\neq j$ the following auxiliary variables:
      $$S_{ij}=(x_i -x_j)(x_i-x_j)+(y_i-y_j)(y_i-y_j), \hspace{0.5 cm} SS_{ij}={S_{ij}}^{-3/2} $$
Then from (2) we have:
     $$\dot{x}_i={vx}_i$$
     $$\dot{y}_i={vy}_i$$
     $$\dot{vx}_i=\sum_{j=1,j\neq i}^{3}(x_j-x_i)SS_{ij}$$
     $$\dot{vy}_i=\sum_{j=1,j\neq i}^{3}(y_j-y_i)SS_{ij}$$
Obviously $S_{ij}=S_{ji}$ and $SS_{ij}=SS_{ji}$, so the memory and computations for them
can be reduced by half. We have the following formulas for the normalized derivatives
for $S_{ij}, SS_{ij}, x_i, y_i, vx_i, vy_i$.
$$S_{ij}^{[k]}=\sum_{m=0}^{k}({x_i}^{[k-m]}-{x_j}^{[k-m]})({x_i}^{[m]}-{x_j}^{[m]})+({y_i}^{[k-m]} -{y_j}^{[k-m]})({y_i}^{[m]}-{y_j}^{[m]}), k=0,1,...$$
$$SS_{ij}^{[0]} = {(S_{ij}^{[0]})}^{-3/2}, \hspace{0.2 cm} SS_{ij}^{[k]} = \frac{1}{2kS^{[0]}}\sum_{m=0}^{k-1}(m-3k)S_{ij}^{[k-m]}SS_{ij}^{[m]}, \hspace{0.2 cm} k=1,2,3,...$$
$$x_i^{[k+1]}=vx_i^{[k]}/(k+1), \hspace{0.2 cm} y_i^{[k+1]}=vy_i^{[k]}/(k+1), \hspace{0.2 cm} k=0,1,...$$
$$vx_i^{[k+1]}=\bigg( \sum_{j=1 j\neq i}^{3}\sum_{m=0}^{k}({x_j}^{[k-m]}-{x_i}^{[k-m]})SS_{ij}^{[m]}\bigg)/(k+1), \hspace{0.2 cm} k=0,1,...$$
$$vy_i^{[k+1]}=\bigg( \sum_{j=1 j\neq i}^{3}\sum_{m=0}^{k}({y_j}^{[k-m]}-{y_i}^{[k-m]})SS_{ij}^{[m]}\bigg)/(k+1), \hspace{0.2 cm} k=0,1,...$$
We need additional formulas for computing the normalized derivatives for the partial derivatives with respect to $v_x$ and $v_y$.
Let $p$ be a general notion for the parameters $v_x$, $v_y$. We need auxiliary variables $\frac{\partial S_{ij}}{\partial p}$, $\frac{\partial SS_{ij}}{\partial p}.$
The formula for $\frac{\partial S_{ij}^{[k]}}{\partial p}$ is straightforward:
$$\frac{\partial S_{ij}^{[k]}}{\partial p}=2\sum_{m=0}^{k}( {x_i}^{[k-m]} -{x_j}^{[k-m]})(\frac{\partial {x_i}^{[m]}}{\partial p}-\frac{\partial{x_j}^{[m]}}{\partial p})+({y_i}^{[k-m]} -{y_j}^{[k-m]})(\frac{\partial{y_i}^{[m]}}{\partial p}-\frac{\partial{y_j}^{[m]}}{\partial p}),$$
$$k=0,1,...$$
The most difficult formula is for $\frac{\partial SS_{ij}^{[k]}}{\partial p}.$
Automatic differentiation for partial derivatives in the case of power of a function is considered in detail in \cite{Barrio2}.
For completeness we derive the formula for $\frac{\partial SS_{ij}^{[k]}}{\partial p}$ here.
Omitting the index $ij$ for brevity we have:
For k=0: $$\frac{\partial SS^{[0]}}{\partial p}=-\frac{3}{2}SS^{[0]}\frac{\partial S^{[0]}}{\partial p}/S^{[0]}$$
For $k>=1$:
$$ SS^{[k]}S^{[0]}=\frac{1}{2k}\sum_{m=0}^{k-1}(m-3k)SS^{[m]}S^{[k-m]}$$
$$\frac{\partial SS^{[k]}}{\partial p}S^{[0]}+SS^{[k]}\frac{\partial S^{[0]}}{\partial p}=\frac{1}{2k}\sum_{m=0}^{k-1}(m-3k)(\frac{\partial SS^{[m]}}{\partial p}S^{[k-m]} + SS^{[m]}\frac{\partial S^{[k-m]}}{\partial p})$$
$$\frac{\partial SS^{[k]}}{\partial p}=\frac{1}{S^{[0]}}\bigg(\frac{1}{2k}\sum_{m=0}^{k-1}(m-3k)(\frac{\partial SS^{[m]}}{\partial p}S^{[k-m]} + SS^{[m]}\frac{\partial S^{[k-m]}}{\partial p})-SS^{[k]}\frac{\partial S^{[0]}}{\partial p}\bigg)$$
At last we have:
$$\frac{\partial x_i^{[k+1]}}{\partial p}=\frac{\partial vx_i^{[k]}}{\partial p}/(k+1), \hspace{0.2 cm} \frac{\partial y_i^{[k+1]}}{\partial p}=\frac{\partial {vy_i^{[k]}}}{\partial p}/(k+1), \hspace{0.2 cm}  k=0,1,...$$

$$\frac{\partial vx_i^{[k+1]}}{\partial p}=\bigg(\sum_{j=1 j\neq i}^{3}\sum_{m=0}^{k}(\frac{\partial {x_j}^{[k-m]}}{\partial p}-\frac{\partial {x_i}^{[k-m]}}{\partial p})SS_{ij}^{[m]}
+({x_j}^{[k-m]}-{x_i}^{[k-m]})\frac{\partial SS_{ij}^{[m]}}{\partial p}\bigg)/(k+1),$$  $$k=0,1,...$$
$$\frac{\partial vy_i^{[k+1]}}{\partial p}= \bigg( \sum_{j=1 j\neq i}^{3}\sum_{m=0}^{k}(\frac{\partial {y_j}^{[k-m]}}{\partial p}-\frac{\partial {y_i}^{[k-m]}}{\partial p})SS_{ij}^{[m]}
+({y_j}^{[k-m]}-{y_i}^{[k-m]})\frac{\partial SS_{ij}^{[m]}}{\partial p}\bigg)/(k+1),$$  $$k=0,1,...$$

\section{Numerical results}
We apply the numerical procedure explained in previous sections for searching periodic orbits with periods $T\leq50$.
At stage (I) of the numerical procedure, when searching for candidates for the correction method, we use 90 order Taylor series method with 96 decimal digits
of precision. At stage (II) of correction with  the modified Newton's method we use 200 order Taylor series method with
192 decimal digits of precision. The GMP library (the GNU Multiple Precision arithmetic library) \cite{gnu} is used for floating point arithmetic in our C-programs.

A topological method from \cite{Mont} is applied to classify the periodic orbits into families.
Each family corresponds to a different conjugacy  class of the free group on two letters (a,b) \cite{Shuv1}, \cite{Shuv2}.
We use "the free group word reading algorithm" from \cite{Shuv2} to obtain the free group elements.
For each found solution we compute the free group element and the four numbers $(v_x,v_y,T,T^*)$ with 150 correct digits,
where $T^*$ is the scale-invariant period. The scale-invariant period is defined as  $T^{*}=T{|E|}^{\frac{3}{2}},$
where  E is the energy of our initial configuration: $E=-2.5 + 3({v_x}^2+{v_y}^2)$.

We succeed to find 33 new families for $T\leq50$ that are not included in the data base of 695 families in \cite{Liao1}
and also are not their satellites. We attribute the finding of these new families of the simultaneous usage of both a relaxed condition for the return proximity
and the larger domain of convergence of the modified Newton's method. For 16 of these 33 solution the classic Newton's method diverges (fails to converge).
Although we can not compensate the usage of a coarser search grid for $v_x$ and $v_y$ with the usage of a relaxed condition for the return proximity and the modified Newton's
method, it is clear that for a fixed search grid, we obtain more solutions.

For some families we found more than one member - members with very close, but different $T^{*}.$
For each family we pick one representative and order all 33 representatives by increasing $T^*$.
The triplet  $(v_x,v_y,T)$ with 20 correct digits and $T^*$ with 10 correct digits can be seen in Table 1.
The free group elements (words) and their lengths $L_f$ can be seen in Table 2.
The four numbers $(v_x,v_y,T,T^*)$ with 150 correct digits, the free group elements and animations
in the real space for these 33 representatives can be found in \cite{rada3body}.
Let us mention that the Newton's method or its modification allows us to obtain periodic solutions with arbitrary accuracy.
If a higher accuracy is needed, we have to increase the floating point precision and the order of the Taylor series method.
We made additional numerical tests for the stability of the presented solutions.
These tests show that all presented 33 solutions are most possibly unstable.

The computations are performed in "Nestum" cluster, Sofia, Bulgaria \cite{nestum}
and "Govorun" supercomputer, Dubna, Russia \cite{HybriLIT}.

 \begin{table}
 %\begin{tabular}{c c c c c}
 \begin{tabular}{ p{0.4cm} p{3.05cm} p{3.05cm} p{3.05cm} p{1.55cm}}
 \hline
 $N$ & $\hspace{1.3cm}v_x$ & $\hspace{1.3cm}v_y$ & $\hspace{1.3cm}T$ & $\hspace{0.6cm}T^*$ \\
 \hline
   \scriptsize{1} & \scriptsize{0.70019547131736421109e0} & \scriptsize{0.40717185305210581416e0} & \scriptsize{0.45872198143326118451e2} & \scriptsize{0.1779045014e2}  \\
    \scriptsize{2} & \scriptsize{0.55588616644229657156e-1} & \scriptsize{0.47637950140930837551e0} & \scriptsize{0.98706000840467862867e1} & \scriptsize{0.2403431228e2}  \\
    \scriptsize{3} & \scriptsize{0.83258051546064251514e-1} & \scriptsize{0.45974114031032641375e0} & \scriptsize{0.12142233475610030906e2} & \scriptsize{0.3043231491e2}  \\
     \scriptsize{4} & \scriptsize{0.52468103235623716206e0} & \scriptsize{0.55808949874280643065e0} & \scriptsize{0.48104413683163137708e2} & \scriptsize{0.3060565106e2}  \\
   \scriptsize{5} & \scriptsize{0.16415866506178833904e0} & \scriptsize{0.72375698700003705153e0} & \scriptsize{0.42370888555129063843e2} & \scriptsize{0.33068766e2}  \\
    \scriptsize{6} & \scriptsize{0.43166431805312424788e-1} & \scriptsize{0.50231939957460362542e0} & \scriptsize{0.1839408945255045597e2} & \scriptsize{0.4212514251e2}  \\
    \scriptsize{7} & \scriptsize{0.49572694266921606895e-1} & \scriptsize{0.3729295241648047833e0} & \scriptsize{0.14879463624676161661e2} & \scriptsize{0.4448765129e2}  \\
   \scriptsize{8} & \scriptsize{0.24076986401710818951e-1} & \scriptsize{0.46576657385237836335e0} & \scriptsize{0.18713457591687910967e2} & \scriptsize{0.4699065001e2}  \\
    \scriptsize{9} & \scriptsize{0.10105579859694620219e0} & \scriptsize{0.48832513783818875985e0} & \scriptsize{0.2484137310892532305e2} & \scriptsize{0.5770482501e2}  \\
   \scriptsize{10} & \scriptsize{0.27326678827008290676e0} & \scriptsize{0.45414874698828708882e0} & \scriptsize{0.29697794877533828243e2} & \scriptsize{0.6335712303e2}  \\
    \scriptsize{11} & \scriptsize{0.63330864579878416347e-1} & \scriptsize{0.27280725486231833479e0} & \scriptsize{0.18976175171090612557e2} & \scriptsize{0.6467308987e2}  \\
    \scriptsize{12} & \scriptsize{0.63381042333416025288e-1} & \scriptsize{0.37416937515175178766e0} & \scriptsize{0.23154023046648329116e2} & \scriptsize{0.6885468209e2}  \\
   \scriptsize{13} &  \scriptsize{0.49215445147391167364e-1} & \scriptsize{0.27208783274830359935e0} & \scriptsize{0.20146498376921738155e2} & \scriptsize{0.6893209371e2}  \\
    \scriptsize{14} & \scriptsize{0.17607268420642445286e0} & \scriptsize{0.27212704370395392306e0} & \scriptsize{0.21486498889242167674e2} & \scriptsize{0.6938951853e2}  \\
   \scriptsize{15} & \scriptsize{0.15788098439573786015e0} & \scriptsize{0.27109743325402938352e0} & \scriptsize{0.22505858851452329866e2} & \scriptsize{0.7367692044e2}  \\
   \scriptsize{16} & \scriptsize{0.66924285481485110315e-1} & \scriptsize{0.30556400601988133149e0} & \scriptsize{0.23855893548806558981e2} & \scriptsize{0.7818769296e2}  \\
   \scriptsize{17} & \scriptsize{0.11369309331730877353e0} & \scriptsize{0.1012774655390907586e0} & \scriptsize{0.20914731627496395039e2} & \scriptsize{0.7924692223e2}  \\
   \scriptsize{18} & \scriptsize{0.77068420570266888116e-1} & \scriptsize{0.87555133287629918407e-1} & \scriptsize{0.2058829098082364828e2} & \scriptsize{0.7939748528e2}  \\
    \scriptsize{19} & \scriptsize{0.34063368427518438244e-1} & \scriptsize{0.78160130410592008104e-1} & \scriptsize{0.21442631372528985026e2} & \scriptsize{0.8365280669e2}  \\
   \scriptsize{20} & \scriptsize{0.65157560764305407762e-1} & \scriptsize{0.26489894552983379382e0} & \scriptsize{0.24648449240704021366e2} & \scriptsize{0.8467639647e2}  \\
   \scriptsize{21} & \scriptsize{0.2678881760018329646e0} & \scriptsize{0.31325597946443836973e0} & \scriptsize{0.31627812537686632459e2} & \scriptsize{0.8880828173e2}  \\
    \scriptsize{22} & \scriptsize{0.17438294015564519865e0} & \scriptsize{0.26762680769678411049e0} & \scriptsize{0.27448264355152555561e2} & \scriptsize{0.891948882e2}  \\
   \scriptsize{23} & \scriptsize{0.11505165723913174772e0} & \scriptsize{0.12491730801530585699e0} & \scriptsize{0.26430807947250193098e2} & \scriptsize{0.9910030917e2}  \\
    \scriptsize{24} & \scriptsize{0.1985931202673251025e0} & \scriptsize{0.10471731921249237968e0} & \scriptsize{0.2757952649739498953e2} & \scriptsize{0.9927770957e2}  \\
   \scriptsize{25} & \scriptsize{0.17668018232516138223e0} & \scriptsize{0.24223620895976353032e0} & \scriptsize{0.29810813421102310101e2} & \scriptsize{0.9929417595e2}  \\
    \scriptsize{26}& \scriptsize{0.18740989368587356788e0} & \scriptsize{0.69789057278404458052e-1} & \scriptsize{0.27044095912662950994e2} & \scriptsize{0.9929872379e2}  \\
    \scriptsize{27} & \scriptsize{0.77563603835657024936e-1} & \scriptsize{0.31659601895264757976e0} & \scriptsize{0.33462918207618659313e2} & \scriptsize{0.1078009873e3}  \\
   \scriptsize{28} & \scriptsize{0.13220711892514996964e0} & \scriptsize{0.45303445850879457837e0} & \scriptsize{0.44311100697379183573e2} & \scriptsize{0.1098613605e3}  \\
    \scriptsize{29} & \scriptsize{0.13268520792396480936e0} & \scriptsize{0.45602157532501814319e0} & \scriptsize{0.47368843757714037722e2} & \scriptsize{0.116623484e3}  \\
   \scriptsize{30} & \scriptsize{0.11713138950134826252e0} & \scriptsize{0.1356481768604369396e0} & \scriptsize{0.31932240890396116136e2} & \scriptsize{0.1189962776e3}  \\
   \scriptsize{31} & \scriptsize{0.12314756779310208734e0} & \scriptsize{0.12031996986120324751e0} & \scriptsize{0.31845635558272641952e2} & \scriptsize{0.1192245119e3}  \\
    \scriptsize{32} & \scriptsize{0.26227767215284254412e0} & \scriptsize{0.3393499442650385466e0} & \scriptsize{0.47045772572847176958e2} & \scriptsize{0.1279252394e3}  \\
   \scriptsize{33} & \scriptsize{0.24818940997429134275e0} & \scriptsize{0.28500096891163700291e0} & \scriptsize{0.43241385269952612616e2} & \scriptsize{0.1289247353e3}  \\

\hline
\end{tabular}

{\caption {Initial conditions $v_x, v_y$, period $T$, scale-invariant period $T^*$ for all 33 solutions}}
\label{tab1}
\end{table}

 \begin{table}
 %\begin{tabular}{c c c c c}
 \begin{tabular}{ p{0.4cm} p{10.2cm} p{0.8cm} }
 \hline
 $N$ & free group element & $L_f$ \\
 \hline
    \scriptsize{1} & bAAbaBBa &  8 \\
    \scriptsize{2} & bABaBabAbABa &  12 \\
    \scriptsize{3} & bABaBAbaBAbABa &  14 \\
    \scriptsize{4} & bAbabAbaBabaBa &  14 \\
    \scriptsize{5} & bABabbABaabABa &  14 \\
    \scriptsize{6} & bABabABBabAABabABa &  18 \\
    \scriptsize{7} & bABBAbabABabaBAABa &  18 \\
    \scriptsize{8} & bABaBabABBabAABabAbABa &  22 \\
    \scriptsize{9} & bABaBAbaBAbbABaaBAbaBAbABa &  26 \\
    \scriptsize{10} & bABAbaBABabaBAbabABAbaBABa &  26 \\
    \scriptsize{11} & baBABabaBABAbaBABAbabABAba &  26 \\
    \scriptsize{12} & bABBAbabABabaBAbabABabaBAABa &  28 \\
    \scriptsize{13} & baBABabaBAbabABabaBAbabABAba &  28 \\
    \scriptsize{14} & baBABababABaAbaBbABababABAba &  28 \\
    \scriptsize{15} & baBABababABAAAbaBBBABababABAba &  30 \\
    \scriptsize{16} & baBABabaBAbaBABabABAbaBAbabABAba &  32 \\
    \scriptsize{17} & baBAABabbaBAABababABBAbaabABBAba &  32 \\
    \scriptsize{18} & baBAABAbabaBABababABAbabaBABBAba &  32 \\
    \scriptsize{19} & baBAAABabbaBABAbabaBABAbaabABBBAba &  34 \\
    \scriptsize{20} & baBABababABAbabaBAbabaBABababABAba &  34 \\
    \scriptsize{21} & baBABabABABabaBABabABAbabABABabABAba &  36 \\
    \scriptsize{22} & baBABababABAAbaBbABaAbaBBABababABAba &  36 \\
    \scriptsize{23} & baBAABabbaBABAbaabABABabbaBABAbaabABBAba &  40 \\
    \scriptsize{24} & baBABAbaabABBAbabaBABAbabaBAABabbaBABAba &  40 \\
    \scriptsize{25} & baBABababABAABabbaBABAbaabABBABababABAba &  40 \\
    \scriptsize{26} & baBABAbaabaBABAbabABABabaBABAbabbaBABAba &  40 \\
    \scriptsize{27} & baBABabABAbaBABabABAbabaBABabABAbaBABabABAba &  44 \\
    \scriptsize{28} & bABAbaBAbaBAbabABabABAbaBABabABabaBAbaBAbaBABa &  46 \\
    \scriptsize{29} & bABabaBAbaBAbaBAbaBAbaBAbaBAbaBAbaBAbaBAbaBAbabABa &  50 \\
    \scriptsize{30} & baBAABabbaBABAbabaBAABababABBAbabaBABAbaabABBAba & 48 \\
    \scriptsize{31} & baBAABabbaBABAbabaBABAbabaBABAbabaBABAbaabABBAba & 48 \\
    \scriptsize{32} & baBABabABAbabABAbabABAbabABabaBABabaBABabaBABabABAba  & 52 \\
    \scriptsize{33} & baBABabaBABababABAbabABABabABABabaBABababABAbabABAba  & 52 \\

\hline
\end{tabular}

{\caption {Free group element (word) and its length $L_f$ for all 33 solutions }}
\label{tab1}
\end{table}

\section{Conclusions}

A detailed description of the classic Newton's method and its modification
based on the Continuous analog of  Newton's method for computing periodic orbits of the planar three-body problem is presented.
Using the formulas from the paper, one can easily write their own programs for searching periodic orbits
for the planar three-body problem. The numerical results show that relaxing the
condition for the return proximity and using the modified Newton's method can help to find
more solutions for a given search grid for $v_x$ and $v_y$.

\newpage

\begin{acknowledgement}

We thank for the opportunity to use the computational resources of the Nestum cluster, Sofia, Bulgaria.
We also thank the Meshcheryakov Laboratory of Information Technologies of JINR, Dubna, Russia for the opportunity to use the computational resources of the HybriLIT Heterogeneous Platform.
The work is supported by a grant of the Plenipotentiary Representative of the Republic of Bulgaria at JINR, Dubna, Russia.

\end{acknowledgement}

\end{document}